\documentstyle[amssymb,12pt]{article}

\newtheorem{definition}{Definition}
\newtheorem{proposition}{Proposition}
\newtheorem{properties}{\sc Property}
\newtheorem{corollary}{Corollary}
\newtheorem{exemple}{\sc Exemple}

\def\be{\begin{equation}}
\def\ee{\end{equation}}
\def\bea{\begin{eqnarray}}
\def\eea{\end{eqnarray}}

\def\bdf{\begin{definition}}
\def\edf{\end{definition}}
\def\bpr{\begin{properties}}
\def\epr{\end{properties}}
\def\bpt{\begin{proposition}}
\def\ept{\end{proposition}}
\def\bcll{\begin{corollary}}
\def\ecll{\end{corollary}}
\def\bex{\begin{exemple}}
\def\eex{\end{exemple}}

\begin{document}

\title{Answer to the Comment about the Letter entitled ``Scalar fields as dark
matter in spiral galaxies'' }
\author{F. Siddhartha Guzm\'an and Tonatiuh Matos \\
{\small {\it Departamento de F\'{\i}sica}}\\
{\small {\it Centro de Investigaci\'on y de Estudios Avanzandos del IPN}}\\
{\small {\it Apdo. Postal 14-740 M\'{e}xico \ 07000 D.F., M\'{e}xico}}}
\maketitle

\renewcommand{\thefootnote}{\fnsymbol{footnote}}


\bigskip

The comment about reference \cite{Ton} starts with false arguments and
therefore the conclusions of the comment are wrong. Nevertheless, the
arguments used by the author are often motive of confusions and it is worth
to publish this answer in order to avoid misunderstandings in the
future.\newline

It is well known that there are no non-singular asymptotically flat, static
solutions to the coupled Einstein Scalar Field theory. However, as it is
discussed in \cite{largo}, the space-time of a galaxy cannot be
asymptotically flat, otherwise, the rotation curves of the galaxy would
decay. Observations show that rotation curves in galaxies grow up for larger
radii or remain flat, but in general they do not decay. The ``source'' of
the scalar field should therefore provoke a non-asymptotically flat
space-time as well. The space-time of \ the paper \cite{Ton} is singular
only for $r=0$, but as it is specified in it, it pretends to be only the
metric of the dark matter dominated region. The study of the center of the
galaxy is much more complicated because we do not have any direct
observation of it. If we follow the hypothesis of the scalar dark matter, we
cannot expect that the center of a galaxy is made of \ ``ordinary'' matter,
we expect that it contains baryons, self-interacting scalar fields, etc.
There the density contrast of baryonic matter and scalar dark matter is the
same (see \cite{cdm}) and their states are in extreme conditions. As long as
we know, there are no existence-uniqueness theorems for non-asymptotically
flat, axial symmetric space-times under these conditions.\\

Furthermore, the energy conditions are no longer valid in nature, as it is
shown by cosmological observations on the dark energy. Why should the energy
conditions be valid in a so extreme state of matter like it is observed in
the center of galaxies? If dark matter is of scalar nature, why should it
fulfill the energy conditions in such extreme situation? As long as we know,
there is no uniqueness theorem in the presence of matter like in this
situation either.\newline

The second objection of the author is related with the geodesics of the
metric. In fact, this point is not clear enough in the paper \cite{Ton}, but
its statement is not new, it was discussed in \cite{AnnPhys} and maybe it
could be the reason to publish the comment. In \cite{largo} we found
that if the dark matter is scalar, either the scalar potential vanishes or
the velocities of the stars are luminal. This is equivalent to the statement
of the author of the comment. However, even when the luminous matter only
represents about 5-10\% of the whole matter of the galaxy, it is crucial for
its stability. The geodesic equations of the metric (21) of reference\cite
{Ton} (representing only the dark matter component) in the equatorial plane
read

\begin{equation}
\frac{d^{2}D}{d\tau ^{2}}-D\left( \frac{d\phi }{d\tau }\right)
^{2}+f_{0}^{2}c^{2}D\left( \frac{dt}{d\tau }\right) ^{2}=0,\ \ \ \frac{d\phi 
}{d\tau }=\frac{B}{D^{2}f_{0}},\ \ \ \frac{dt}{d\tau }=\frac{A}{%
c^{2}D^{2}f_{0}}  \label{geo}
\end{equation}

\noindent where $\tau $ is the proper time of the test particle and $D=\int {%
ds}=\sqrt{(r^{2}+b^{2})/f_{0}}$ is the proper distance of the test particle
at the equator from the galactic center (we set $a=0,$ and $r_{0}=1$).
Observe that for a circular trajectory it follows

\begin{equation}
\frac{d\phi }{dt}=f_{0}c=c^{2}\frac{B}{A}
\end{equation}

\noindent Moreover $B=f_{0}cD$ along the whole galaxy. The first of
equations (\ref{geo}) is the second Newton's law for particles travelling
onto the scalar field background. We can interpret

\begin{equation}
\frac{d^{2}D}{d\tau ^{2}}=D\left( \frac{d\phi }{d\tau }\right)
^{2}-f_{0}^{2}c^{2}D\left( \frac{dt}{d\tau }\right) ^{2}=\frac{B^{2}}{%
D^{3}f_{0}^{2}}-\frac{A^{2}}{c^{2}D^{3}}=\frac{c^{2}}{D}-\frac{A^{2}}{%
c^{2}D^{3}}  \label{chr5}
\end{equation}
as the force due to the scalar field background, i.e. $F_{\Phi
}=c^{2}/D-A^{2}/\left( c^{2}D^{3}\right) $. Using the expression for $A$
given in equation (16) of \cite{Ton}, we can write this force in terms of $v$

\[
F_{\Phi }=-\frac{v^{2}}{D\left( f_{0}^{2}D^{2}-v^{2}/c^{2}\right) }
\]
which corresponding potential is

\begin{equation}
V_{\Phi }=\frac{1}{2}c^{2}\ln \left( f_{0}^{2}-\frac{1}{D^{2}}\frac{v^{2}}{%
c^{2}}\right)   \label{Vp}
\end{equation}
Of course, as the author of the comment pointed out, this potential
corresponds to non stable trayectories. Nevertheless, $v^{2}/c^{2}\simeq
10^{-6}$, and $f_{0}\simeq 0.01$ $kpc^{-1}$, this means that potential $%
V_{\Phi }$ is almost constant for $D\gtrsim 0.3$ $kpc$, which corresponds to
the region where the solution is valid. At the other hand, we know that the
luminous matter is completely Newtonian. The Newtonian force due to the
luminous matter is given by $F_{L}=GM(D)/D^{2}=v_{L}^{2}/D=B_{L}^{2}/D^{3}$,
where $v_{L}$ is the circular velocity of the test particle due to the
contribution of the luminous matter, given by equation (23) in \cite{Ton}
and $B_{L}=B_{L}(D)$ is its corresponding angular momentum per unit of mass,
given by equation (24) in \cite{Ton}. The total force acting on the test
particle is then $F=F_{L}+F_{\Phi }.$ For circular trajectories $%
d^{2}D/d\tau ^{2}=F=0$, then

\begin{equation}
F=\frac{B_{L}^{2}}{D^{3}}-\frac{A^{2}}{c^{2}D^{3}}+\frac{c^{2}}{D}=0
\label{chr6}
\end{equation}
which is just equation (18) of reference \cite{Ton}. The corresponding
potential is then $V=V_{L}+V_{\Phi }$, but for the regions where $D\gtrsim
0.3$ $kpc$, potential $V$\ is dominated by the behavior of $V_{L}$, which
contains stable circular trayectories, like in a galaxy. Therefore the argument
of the author of the comment is false, he is not considering the luminous
matter in his analysis. Of course, without luminous matter (without a
galaxy) the system is unstable.\newline
For regions with $D\lesssim 0.3$ $kpc$ the solution in not valid any more
due to the approximations we have carried out.\\

Therefore the last conclusion of the comment is completely wrong, the
hypothesis of the scalar dark matter is well justified at galactic level
(see also \cite{sph}) and at the cosmological level too \cite
{cdm,cosmos1,cosmos2}. But at this moment it is only that, a hypothesis
which is worth to be investigated.\newline

\end{document}